# Modeling the Distribution of the Arrival Angle Based on Transmitter Antenna Pattern

Cezary Ziółkowski, Jan M. Kelner, Leszek Nowosielski, Marian Wnuk
Institute of Telecommunications, Faculty of Electronics, Military University of Technology, Warsaw, Poland
{cezary.ziolkowski, jan.kelner, leszek.nowosielski, marian.wnuk}@wat.edu.pl

*Abstract*—An angular distribution of received signals has a significant impact on their spectral and correlational properties. Most of angular dispersion models do not consider antenna patterns. The developed procedure for determining the propagation path parameters enables a wide range of assessment of the impact of the propagation environment on the received signal properties. In contrast to the other models, this procedure is based on a geometrical structure, which parameters are defined on the basis of power delay profile or spectrum This modeling method allows also the power radiation pattern (PRP) of the transmitting antenna. The aim of the paper is to present the influence of the transmitter antenna PRP on the scattering propagation paths that arrive at the receiver. This analysis is realized on the basis of simulations studies using the developed procedure. Presented in this paper procedure maps the effects of propagation phenomena that predominate in an azimuth plane.

## I. Introduction

Distribution of the arrival angle has a significant impact on the spectral and correlational properties of the received signals and determines the methods of signal processing. Evaluation of the these methods' effectiveness is based primarily on simulation tests, which should include a wide range of factors that determine the propagation phenomenon. Developed procedure ensures generation of a set of angles and powers of propagation paths arriving at the receiver (Rx). This procedure considers the impact of the transmitting antenna pattern on the intensity of the angular dispersion of the received signals. In the literature, only simplified modeling methods of the impact of the antenna patterns are presented. Mostly, these methods are based on the assumption that transmitting antenna has omnidirectional pattern. Presented in this paper procedure maps the effects of propagation phenomena that predominate in the azimuth plane. In practice, this means that the propagation phenomena analysis refers to the transmitting antenna, the power radiation pattern (PRP) in the elevation plane is narrow (i.e. few/several degrees).

The aim of the paper is to present the modeling of the transmitter antenna PRP on the scattering propagation paths that arrive at Rx. In this paper, the influence of the PRP width on the rms angle spread (AS) is presented. This analysis is based on parameters of the real channels. In the simulation tests, the Gaussian model is adopted to describe the PRP of the transmitting antenna in the azimuth plane. Comparison of the simulation results with measurement data is the basis for verification of the modeling procedure the intensity of the angular dispersion of the received signals.

The paper is organized as follows. In Section II, the basic assumptions of the procedures are presented. Section III describes the method for determining the set of angles and powers for the delayed and local scattering components of the received signal. The verification of the procedure is performed on the basis of the measurement data, the results of which are presented in Section IV. The assessment of the influence of the width of the transmitting antenna radiation pattern on AS of the signals is shown in Section V. Summary is contained in Section VI.

## II. Fundamentals of Modeling Procedure

The developed procedure uses geometric and statistical modeling methods of propagation. For the delayed scattering components, the multi-elliptical model [1] is adopted and the von Mises distribution is used to the description of the angular dispersion of the local scattering components. In Fig. 1, the geometry of the scatterer positions is shown for the delayed components.

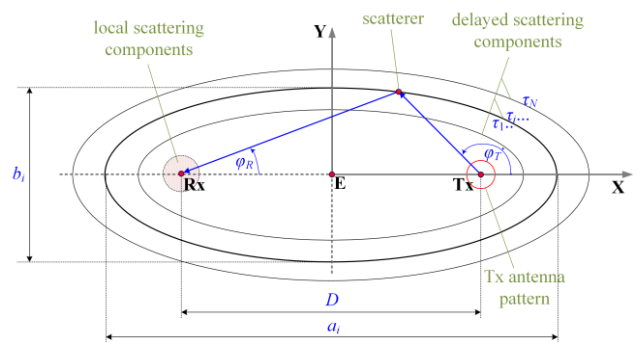

Fig. 1. Geometry of the scatterer positions for delayed components.

The number of local extremes of the power delay profile (PDP) or power delay spectrum (PDS) and the corresponding them the arguments are the basis for the parameters of each ellipse

$$2a_i = D + c\tau_i, \quad e_i = D/2a_i \qquad (1)$$

where $\tau_i$ is the argument of the $i$th local extreme of PDP/PDS, $i = 1,2,…,N$, $N$ is the number of the local extremes (number of the ellipses), c is the speed of light, $D$ is the distance between Rx and the transmitter (Tx), $2a_i$ and $e_i$ are the major axis and eccentricity of the $i$th ellipse, respectively.

For the multi-elliptical model, probability density function (PDF) of the angle of arrival (AOA) of the delayed scattering components, $f_d(\varphi_R)$, has the form

$$f_d(\varphi_R) = \sum_{i=1}^{N} \frac{P_i}{P_R} f_{di}(\varphi_R) \qquad (2)$$

where $P_i = \sum_{j=1}^{M_i} p_{ij}$ is the mean power of the delayed scattering components for $\tau_i$, $p_{ij}$ is the power of the $ij$th path, $i = 0,1,…N$, $j = 1,2,…M_i$, $P_R = \sum_{i=0}^{N} P_i$ is the mean power of the received signal, and $f_{di}(\varphi_R)$ is PDF of AOA for the $i$th ellipse.

In the case of the local scattering components, PDF of AOA is described by the relationship [2]

$$f_l(\varphi_R) = \frac{\exp(\mu \cos \varphi_R)}{2\pi I_0(\mu)}, \quad \varphi_R \in \langle -\pi, \pi), \quad \mu \geq 0 \qquad (3)$$

where $I_0(\cdot)$ refers the zero-order modified Bessel function and $\mu$ is constant, that controls AS.

For a general case, when in addition to the scattering components we consider the direct path component, is [3]

$$f(\varphi_R) = \sum_{i=1}^{N} \frac{P_i}{P_R} f_{di}(\varphi_R) + \frac{1}{\kappa+1} \frac{P_0}{P_R} \frac{\exp(\mu \cos \varphi_R)}{2\pi I_0(\mu)} + \frac{\kappa}{\kappa+1} \frac{P_0}{P_R} \delta(\varphi_R) \qquad (4)$$

where $P_0$ refers the power of all components with zero delay, $\kappa$ is the Rician factor, and $\delta(\cdot)$ is the delta distribution.

This relationship enables the assessment of the impact of the transmitter antenna PRP on the scattering intensity of the signal reception angle for different environmental conditions.

### III. PARAMETERS OF ARRIVAL PROPAGATION PATHS

Assumption of the narrow PRP in the elevation plane allows to simplify the procedure to modeling in two dimensions. The following data are the basis for the designation of the set of angles and powers of the received signal components:

- the distance $D$ between Tx and Rx;
- PDS or PDP, which are the basis for determining the delays, $\{\tau_i\}_{i=0,1,…,N}$, and corresponding mean powers, $\{P_i\}_{i=0,1,…,N}$;
- the number of propagation paths $M_i$ that arrive at Rx with delay $\tau_i$, the average power of which is $P_i/M_i$;
- the normalized radiation pattern $g_T(\varphi_T)$ and the half-power beamwidth (HPBW) of the transmitter antenna.

The generation procedure of powers and angles of the delayed components, which is based on the multi-elliptical model, is as follows:

- determination of the ellipses parameters (major half-axis, $a_i$, and eccentricity, $e_i$, of the $i$th ellipse);
- generation of the angles of departure (AODs) of propagation paths;
- determination of AOAs for Rx location.

The assumption that the probability of scattering element occurrence, which are located on the ellipses with the foci in Rx and Tx, is the same in every direction from Tx, is the basis for AOD generation. For the normalized PRP of the transmitting antenna in the azimuth plane, we have:

$$\frac{1}{2\pi} \int_{-\pi}^{\pi} g_T^2(\varphi_T) d\varphi_T \quad \text{and} \quad g_T^2(\varphi_T) \geq 0 \qquad (5)$$

It is the basis for the adoption of PDF AOD, $f_T(\varphi_T)$ as

$$f_T(\varphi_T) = \frac{1}{2\pi} g_T^2(\varphi_T) \quad \text{for} \quad \varphi_T \in \langle -\pi, \pi) \qquad (6)$$

For the Gaussian model of the transmitting antenna PRP [4]

$$g_T(\varphi_T) = C_0 \exp\left(-\frac{\varphi_T^2}{2\sigma_T}\right) \quad (C_0 - \text{normalizing constant}) \qquad (7)$$

we have

$$f_T(\varphi_T) = C(\sigma_T) \exp\left(-\frac{\varphi_T^2}{\sigma_T}\right) \quad \text{for} \quad \varphi_T \in \langle -\pi, \pi) \qquad (8)$$

where $C(\sigma_T) = \left(\int_{-\pi}^{\pi} f_T(\varphi_T) d\varphi_T\right)^{-1} = \left(\sqrt{\mu}\sigma_T \operatorname{erf}(\pi/\sigma_T)\right)^{-1}$ and $\sigma_T = \frac{HPBW}{2\sqrt{\ln 2}} \approx 0.6 HPBW$.

Expression (8) is the basis for the generation of AOD. The relationship between AOD and AOA results from the ellipse properties [5]

$$\cos\varphi_{R_i} = \frac{2e_i + (1+e_i^2)\cos\varphi_T}{1+e_i^2 + 2e_i\cos\varphi_T} \quad (9)$$

hence

$$\varphi_{R_i} = \operatorname{sgn}\varphi_T \arccos\frac{2e_i + (1+e_i^2)\cos\varphi_T}{1+e_i^2 + 2e_i\cos\varphi_T}$$
$$\text{for} \quad \varphi_T \in \langle -\pi, \pi \rangle \quad (10)$$

Expression (10) is the basis for the estimation $f_{di}(\varphi_R)$ for the $i$th ellipse. In the case of the local scattering components, AODs are generated directly on the basis of (3).

Each propagation path i.e. each AOA corresponds to the random power $p_{ij}$. The powers of the delayed scattering components, $p_{ij}$, $i = 1,2,\ldots N$, $j = 1,2,\ldots M_i$, are generated on the basis of the uniform distribution

$$f(p_{ij}) = \begin{cases} M_i/(2P_i) & \text{for} \quad p_{ij} \in \langle 0, 2P_i/M_i \rangle \\ 0 & \text{for} \quad p_{ij} \notin \langle 0, 2P_i/M_i \rangle \end{cases} \quad (11)$$

Note that the mean value of this distribution corresponds to $P_i / M_i$ that determined using the PDP / PDS.

The analogous distribution is used to determine the powers of the local scattering components, $p_{0j}$, $j = 1,2,\ldots M_0$,

$$f(p_{0j}) = \begin{cases} (1+\kappa)M_0/(2P_0) \\ \quad \text{for} \quad p_{0j} \in \langle 0, 2P_0/((1+\kappa)M_0) \rangle \\ 0 \\ \quad \text{for} \quad p_{0j} \notin \langle 0, 2P_0/((1+\kappa)M_0) \rangle \end{cases} \quad (12)$$

where $M_0$ is the number of propagation paths that arrive at Rx without delay.

For each AOA, (11) and (12) provide power allocation to individual propagation paths. As a result, we obtain a set of ordered pairs $\{\varphi_{ij}, p_{ij}\}_{\substack{i=0,1,\ldots,N \\ j=1,2,\ldots,M_i}}$. This set is the basis for the evaluation of the angular dispersion of the received signals.

## IV. EVALUATION OF MODELING PROCEDURE

The basis for the assessment of the intensity of angular dispersion of the received signals is PDF of AOA. Based on (4), we know that the estimation of $f(\varphi_R)$ reduces to the estimation $f_{di}(\varphi_R)$ and $f_l(\varphi_R)$. In practice, determination of PDF of AOA reduces to measurement of the power angular spectrum of the received signals. Therefore, we could present the estimators, $\tilde{f}_{di}(\varphi_R)$ and $\tilde{f}_l(\varphi_R)$, in the form of

$$\tilde{f}_{di}(\varphi_R) = \frac{\sum_{j(\varphi_R)}^{M_i(\varphi_R)} p_{ij}}{\sum_{j=1}^{M_i} p_{ij}} \quad \text{and} \quad \tilde{f}_l(\varphi_R) = \frac{\sum_{j(\varphi_R)}^{M_0(\varphi_R)} p_{0j}}{\sum_{j=1}^{M_0} p_{0j}} \quad (13)$$

where $M_i(\varphi_R)$ is the number of propagation paths that arrive at $\varphi_R$ with delay $\tau_i$, $i = 0,1,\ldots,N$.

Based on (4) and (13), it follows that the estimator $\tilde{f}(\varphi_R)$, is

$$\tilde{f}(\varphi_R) = \sum_{i=0}^{N} \frac{\sum_{j=1}^{M_i} p_{ij}}{\sum_{i=0}^{N}\sum_{j=1}^{M_i} p_{ij}} \cdot \frac{\sum_{j(\varphi_R)}^{M_i(\varphi_R)} p_{ij}}{\sum_{j=1}^{M_i} p_{ij}} = \sum_{i=0}^{N} \frac{\sum_{j(\varphi_R)}^{M_i(\varphi_R)} p_{ij}}{\sum_{i=0}^{N}\sum_{j=1}^{M_i} p_{ij}} \quad (14)$$

The evaluation of modeling procedure is based on the measurement data described in [6]. The data used in simulation studies are the same as the [1, Tab. 4]. Due to the omni-directional PRP of the transmitting antenna used in the measurement campaign we have adopted $HPBW = 360º$. The obtained results are averaged over 500 Monte Carlo trials of the modeling procedure. Figures 2 and 3 show the results of the modeling procedure and measurements for the Aarhus and Stockholm scenarios, respectively.

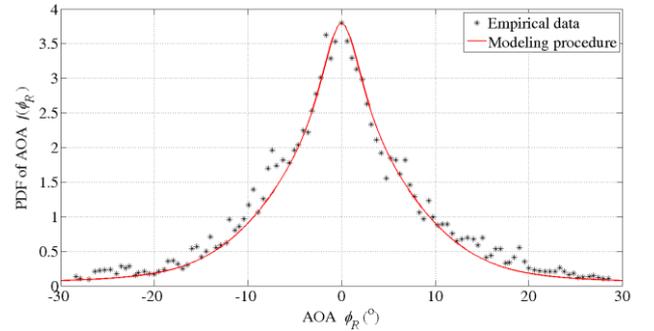

Fig. 2. Simulation and measurement results for Stockholm scenario.

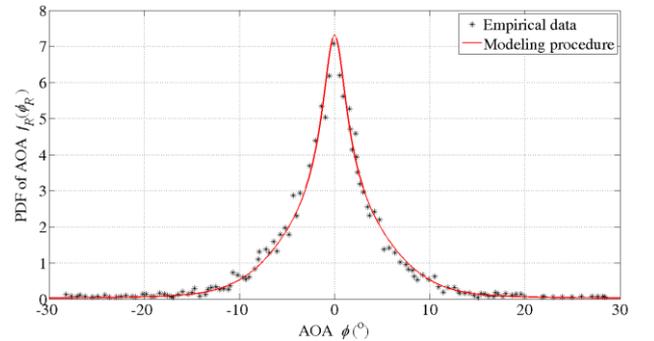

Fig. 3. Simulation and measurement results for Aarhus scenario.

The least-square error (LSE) is the basis for the assessment of the procedure accuracy in comparison to the selected analytical and empirical models that provide the best approximation of the measured data. The results are shown in Table I and Figs. 4 and 5.

TABLE I. COMPARISON OF THE APPROXIMATION RESULTS OF PDF OF AOA

| PDF Model \ Scenario | LSE Stockholm | | LSE Aarhus | |
|---|---|---|---|---|
| Uniform elliptical – Rx outside (UERO) [7] | 0.043075 | ([8]) | 0.084363 | ([8]) |
| Uniform elliptical – Rx inside (UERI) [9],[10],[11] | 0.040088 | ([8]) | 0.196290 | ([8]) |
| Modified Laplacian (ML) [12] | 0.016470 | ([12]) | 0.032001 | ([12]) |
| Modeling procedure | 0.026458 | | 0.039258 | |

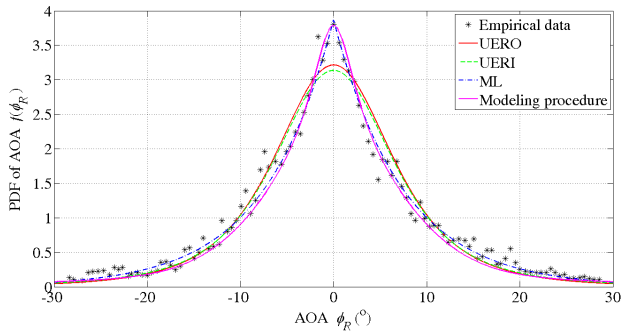

Fig. 4. Fitting of PDF models to empirical data for Stockholm scenario.

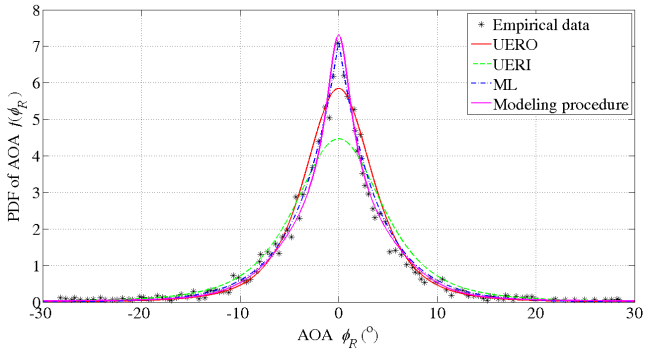

Fig. 5. Fitting of PDF models to empirical data for Aarhus scenario.

The obtained results for the modeling procedure are convergent with the results for the best analytical and empirical models. This is a prerequisite for a positive evaluation of the developed procedure.

## V. ANGULAR DISTRIBUTION VERSUS POWER PATTERN OF TRANSMITTER ANTENNA

The presented procedure of the modeling of the propagation path parameters enables the assessment of the impact of PRP of transmitter antenna on the angular dispersion. Simulation tests are carried out for conditions that result from the measurement scenarios described in [6]. The obtained results show the changes of PDFs of AOA versus the selected HPBW of the transmitting antenna. For the Stockholm and Aarhus scenarios, PDFs of AOA are presented in Figs. 6 and 7, respectively.

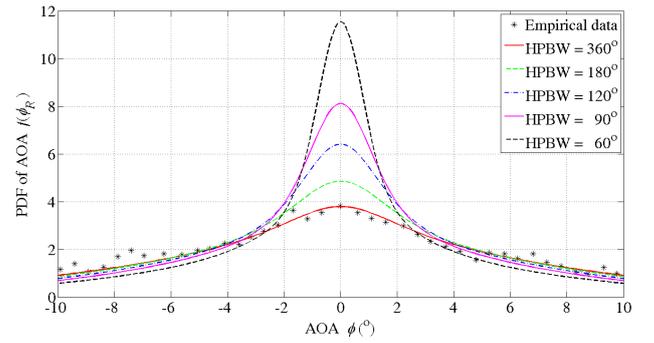

Fig. 6. PDFs of AOA for HPBW = {360º, 180º, 120º, 90º, 60º} – Stockholm scenario.

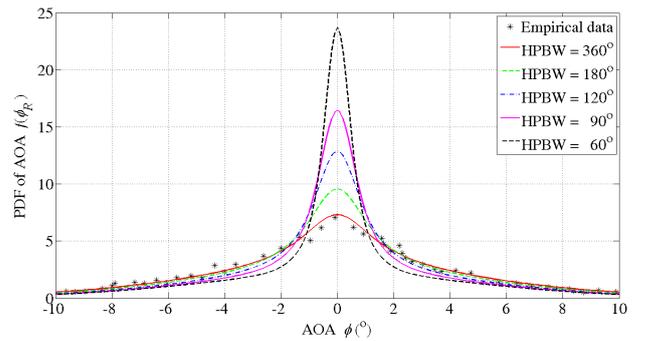

Fig. 7. PDFs of AOA for HPBW = {360º, 180º, 120º, 90º, 60º} – Aarhus scenario.

These figures show that the decreasing of HPBW reduces the angular dispersion of the received signal. As a measure for quantitative evaluation, AS, $\sigma_\varphi$, is used

$$\sigma_\varphi = \sqrt{\sum_{k=1}^{K} \varphi_{Rk}^2 \tilde{f}(\varphi_{Rk}) - \left(\sum_{k=1}^{K} \varphi_{Rk} \tilde{f}(\varphi_{Rk})\right)^2} \qquad (15)$$

where $K$ is the number of all of designated $\varphi_R$ values in simulation study.

For the analyzed scenarios, the relationships between $HPBW$ and $\sigma_\varphi$ are shown in Fig. 8.

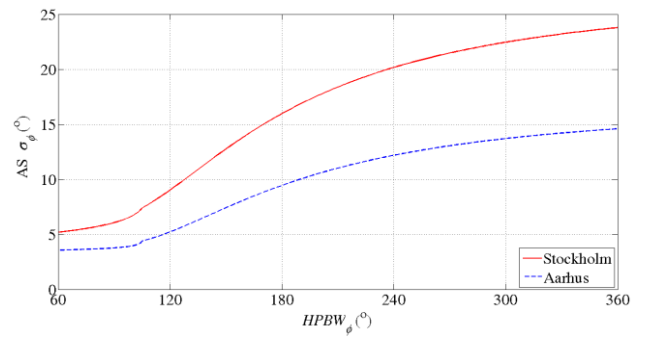

Fig. 8. ASs versus HPBW for Stockholm and Aarhus scenarios.

As shown, the environment for which the rms delay spread, $\sigma_\tau$, is 2.35 µs (Stockholm), introduces a few degrees higher AS in comparison to the environment for which $\sigma_\tau = 1.2$ µs (Aarhus). For both scenarios, Fig. 8 also shows that the largest rate of the AS changes exists for the HPBW changes between 120º and 180º.

## VI. CONCLUSION

The developed procedure for determining the propagation path parameters enables a wide range of assessment of the impact of the propagation environment on the received signal properties. It is based on multi-elliptical model. In contrast to the models that were previously presented in the literature, the developed procedure is based on a geometrical structure, which parameters are defined on the basis of PDP/PDS. The verification of the simulation results with respect to the measurement data indicates the correctness of the developed modeling method of the propagation phenomena effect. In this paper, we focused on the impact of the transmitting antenna PRP on the angular dispersion. The obtained results of simulation tests enable the quantitative assessment of the analyzed issues for different propagation scenarios.